\documentclass[12pt]{iopart}

\usepackage{cite}
\usepackage{graphicx}
\usepackage{multirow}

\begin{document}

\title{The stochastic pump current and the non-adiabatic geometrical phase}

\author{Jun Ohkubo}

\address{
Institute for Solid State Physics, University of Tokyo, 
Kashiwanoha 5-1-5, Kashiwa, Chiba 277-8581, Japan
}
\ead{ohkubo@issp.u-tokyo.ac.jp}
\begin{abstract}
We calculate a pump current in a classical two-state stochastic chemical kinetics
by means of the non-adiabatic geometrical phase interpretation.
The two-state system is attached to two particle reservoirs,
and under a periodic perturbation of the kinetic rates,
it gives rise to a pump current between the two-state system and the absorbing states.
In order to calculate the pump current,
the Floquet theory for the non-adiabatic geometrical phase is extended from a Hermitian case to a non-Hermitian case.
The dependence of the pump current on the frequency of the perturbative kinetic rates
is explicitly derived,
and a stochastic resonance-like behavior is obtained.
\end{abstract}

\maketitle

\section{Introduction}

Some of classical stochastic processes have been investigated 
via analogies with quantum mechanics,
and such analogies give us useful analytical methods for investigating various physical quantities
and phenomena in classical stochastic processes.
For example, asymmetric simple exclusion processes have been studied a lot
by means of the free fermionic representation or the quantum spin formalism\cite{Schutz2000}.
For the reaction-diffusion processes,
it has been shown that the second-quantized representation and field theoretic approach are useful
\cite{Doi1976,Doi1976a,Peliti1985,Mattis1998};
deeper understandings of the non-equilibrium behavior has been achieved
with the aid of the renormalization group method \cite{Tauber2005}.
Recently, the second-quantization method has been applied to
stochastic switching phenomena in gene regulatory networks,
with the aid of a variational method \cite{Sasai2003,Kim2007,Kim2007a,Ohkubo2007}.

Recently, Sinitsyn and Nemenman \cite{Sinitsyn2007,Sinitsyn2007a} 
have been studied a pump current in stochastic chemical kinetics.
They studied a classical two-state stochastic system in a sea of substrates and products (absorbing states);
the system can be interpreted as a single Michaelis-Menten catalyzing enzyme or as a channel on a cell surface.
Under a periodic perturbation of the kinetic rates,
the classical system gives rise to a pump current between the absorbing states.
By means of the expression for the full counting statistics of transitions among the absorbing states,
they have clarified that there is a relationship between the Berry phase and the pump current,
i.e., the pump current can be separated into two parts:
one of them is a classical current, and the other is indeed a geometrical one
which depends only on the contour in the parameter space.
Sinitsyn and Nemenman have calculated the geometrical phase in the adiabatic case,
in which the oscillations of the perturbative kinetic rates are very slow \cite{Sinitsyn2007}.
It has been clarified that
the pump current is linearly proportional to the frequency in the low frequency regime.

In contrast with the low frequency case, such a pump current is expected to show
a power-law decay behavior with respect to the frequency in the high frequency regime 
\cite{Liu1990,Astumian2003,Jain2007}.
In addition, there should be a single maximum in the pump current,
which is a similar behavior to the stochastic resonance.
While this behavior has been obtained by another approach for a system slightly different to 
the Sinitsyn and Nemenman one \cite{Astumian2003},
it is expected to be possible to obtain
the behavior of the pump current by the `geometrical interpretation.'
Furthermore, such studies for classical systems from the viewpoint of the quantum mechanical formulation
might become a key step towards deeper understanding of the classical stochastic systems.

In the present paper,
we analyze the pump current in the `non-adiabatic' case.
In order to evaluate the pump current in the non-adiabatic case,
we will extend the usual Floquet theory, which has been discussed by Moore \cite{Moore1990a}, to a non-Hermitian case.
In addition, the Floquet Hamiltonian cannot be block diagonalized in our case,
and hence a perturbation theory is applied.
From these calculations,
we can derive an explicit dependence of the pump current on the frequency of the perturbative kinetic rates,
which shows a linear dependence on the frequency in the low frequency regime (adiabatic case),
and a power-law decay in the high frequency regime.

The construction of the present paper is as follows.
In section 2, we explain a model for the pump current, which was introduced by Sinitsyn and Nemenman
\cite{Sinitsyn2007}.
The general formalism for the calculation of the pump current is given in section 3.
Section 4 is devoted to an extension of the Floquet theory developed by Moore \cite{Moore1990a} 
to a non-Hermitian case.
In section 5, we exhibit an explicit calculation of the pump current with the aid of perturbation theory.
Section 6 gives concluding remarks.

\section{The model}

Sinitsyn and Nemenman \cite{Sinitsyn2007} have introduced a candidate minimal models 
for the pump current.
The classical chemical kinetic system is constructed as follows:
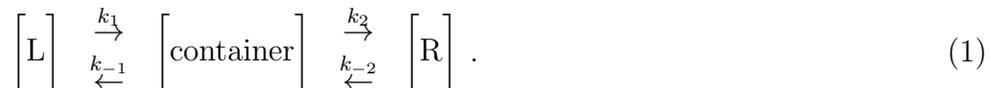
\begin{eqnarray}
\begin{array}{ccccc}
\multirow{2}{20pt}{$\Bigg[\mathrm{L}\Bigg]$}  & \stackrel{k_1}{\rightarrow} & 
\multirow{2}{60pt}{$\Bigg[\mathrm{container}\Bigg]$}  & \stackrel{k_2}{\rightarrow} & 
\multirow{2}{20pt}{$\Bigg[\mathrm{R}\Bigg]$} \\
& \stackrel{k_{-1}}{\leftarrow} &   & \stackrel{k_{-2}}{\leftarrow} & 
\end{array}.
\end{eqnarray}
The system consists of three parts.
The container can contain either zero or one particle in it.
When the container is filled with one particle,
the particle can escape from the container by jumping into one of the two absorbing states:
the left reservoir or the right one.
In contrast, when the container is empty, either of the absorbing states
can emit a new particle into the container.
Here, we define the pump current as the mean particle current $J$ from the container 
into the right reservoir.

For simplicity, 
in the present paper, we consider the following kinetic rates:
\begin{eqnarray}
\begin{array}{l}
k_{-1} = k_2 = a, \\
k_1 = b + R \cos (\omega t), \\
k_{-2} = b + R \sin (\omega t), 
\end{array}
\end{eqnarray}
where $a$ and $b$ are positive real numbers, and $R$ is the amplitude of the perturbative oscillation.
The above kinetic rates mean that 
only two kinetic rates ($k_1$ and $k_{-2}$) oscillate with time at a frequency $\omega$.
For the above kinetic rates, 
there is no classical $L \to R$ current, as discussed in reference \cite{Sinitsyn2007}.
Hence, if there is a current, it is called a pump current, which stems from the geometrical term.

\section{Geometrical phase and pump current}

We here give a brief summary of the discussion given by Sinitsyn and Nemenman \cite{Sinitsyn2007},
and extend their formulation to one useful for the non-adiabatic case.

Let $P_\mathrm{f}$ and $P_\mathrm{e}$ be the probabilities for filled and empty container respectively.
When we define a state of the system by
\begin{eqnarray}
\mathbf{p}(t) = 
\left[ \begin{array}{c}
P_\mathrm{e} \\ P_\mathrm{f}
\end{array}\right],
\end{eqnarray}
the master equation for the time evolution of the system is written as
\begin{eqnarray}
\frac{\rmd}{\rmd t} \mathbf{p}(t)
= - 
\left[ \begin{array}{cc}
k_1 + k_{-2} & - k_{-1} - k_2 \\
-k_1 - k_{-2} & k_{-1} + k_2
\end{array}\right]
\mathbf{p}(t).
\end{eqnarray}
Note that $P_\mathrm{e} = 1 - P_\mathrm{f}$.

In order to investigate the pump current between the container and the right reservoir,
we define $P_n$ as the probability of having $n$ net transitions from the container into the right
reservoir during time $t$.
Using the probability $P_n$, the probability generating function 
for the number of transitions is written by 
\begin{eqnarray}
Z(\chi) = \rme^{S(\chi)} = \sum_{s=-\infty}^{\infty} P_{n=s} \rme^{\rmi s \chi},
\end{eqnarray}
where $\chi$ is called the counting field, and $S(\chi)$ the full counting statistics.
Formally, the generating function $Z(\chi)$ is given by \cite{Sinitsyn2007}
\begin{eqnarray}
Z(\chi) =
\mathbf{1}^\dagger \hat{T} \left( \rme^{-\int_{t_0}^{t} \hat{H} (\chi,t) \rmd t}\right)
\mathbf{p}(t_0),
\end{eqnarray}
where $\mathbf{1}$ is the unit vector, $\hat{T}$ stands for the time-ordering operator, and
\begin{eqnarray}
\hat{H}(\chi,t) = 
\left[ \begin{array}{cc}
k_1 + k_{-2} & - k_{-1} - k_2 \rme^{\rmi \chi} \\
-k_1 - k_{-2} \rme^{- \rmi \chi} & k_{-1} + k_2
\end{array}\right].
\end{eqnarray}

The derivatives of $S(\chi)$ give cumulants of $P_n$,
e.g., $\langle n \rangle = - \rmi \partial S(\chi) / \partial \chi |_{\chi=0}$.
We here denote the period of the rate oscillation by $T (\equiv 2 \pi / \omega)$.
When we assume that a state $\mathbf{p}(t)$ gives a cyclic evolution of the system, 
i.e., $\mathbf{p}(T) = \rme^{K(\chi)} \mathbf{p}(0)$,
the pump current is given by $J_\mathrm{pump} = (-\rmi/T) \partial K(\chi) / \partial \chi |_{\chi=0}$,
because of the normalization condition of the probability ($\mathbf{1}^\dagger \mathbf{p}(0) = 1$).

In order to calculate the full counting statistics (or $K(\chi)$),
we focus on the fact that the probability generating function is related to the solution of the following
differential equation:
\begin{eqnarray}
\frac{\rmd}{\rmd t} \mathbf{p}(t) = - \hat{H}(\chi,t) \mathbf{p}(t).
\end{eqnarray}
When we replace the time-evolution operator $\hat{H}(\chi,t)$ by $H \equiv - \rmi \hat{H}(\chi,t)$,
we obtain the following Shr{\" o}dinger-like equation:
\begin{eqnarray}
\rmi \frac{\rmd}{\rmd t} \phi (t) = H \phi (t)
\label{eq_shrodinger}.
\end{eqnarray}
Hence, the problem of the calculation of the full counting statistics (or $K(\chi)$)
is replaced by the problem of the evaluation of the geometrical phase in equation~\eref{eq_shrodinger}.
We here note that the `Hamiltonian' $H$ is a non-Hermitian operator,
so that we need a slight modification of the usual calculation for the geometrical phase.

\section{Extension of Moore's discussion to a non-Hermitian case}

In order to perform the calculation in the non-adiabatic cases,
one can use the Floquet theory discussed by Moore and Stedman \cite{Moore1990}.
Choutri \textit{et al.} \cite{Choutri2002} have extended the discussion given by Moore and Stedman
to non-Hermitian cases,
and calculated the non-adiabatic geometrical phase for a special case \cite{Choutri2002}.
However, in the above formulations it is difficult to give an explicit expression 
for the geometrical phase for general cases.
On the other hand,
Moore \cite{Moore1990a} have combined the above Floquet theory 
with the discussion given by Shirley \cite{Shirley1965}, 
and obtained a useful formulation for the non-adiabatic geometrical phase in the Hermitian case. 
We will give the extension of the Moore's discussion to non-Hermite cases,
and hence the following discussion is mainly based on the Moore's one \cite{Moore1990a}.

\subsection{The Floquet theorem}

At first, we give a brief summary for the Floquet theorem \cite{Moore1990,Ince1956,Arnold1988}.
A non-singular matrix $A$ is called a \textit{fundamental matrix} of the system of ordinary differential equations
$\dot{x}(t) = W(t) x(t)$ if $\dot{A}(t) = W(t) A(t)$,
where $\dot{x}(t)$ and $\dot{A}(t)$ mean the time derivative of $x(t)$ and $A(t)$, respectively.
Let $W(t)$ be periodic in $\tilde{t}$.
Then, the Floquet theorem ensures that 
there exists a $B(t)$ periodic in $\tilde{t}$ and a constant $C$ such that $A(t) = B(t) \rme^{C t}$.

\subsection{The non-adiabatic geometrical phase and the Floquet theory}

For a periodic non-Hermitian Hamiltonian $H$ with period $T$,
i.e., $H(t+T) = H(t)$,
the non-unitary time-evolution operator $U(t)$
is decomposed into the Floquet product form as $U(t) = V(t) \rme^{\rmi M t}$.
Due to the Floquet theorem,
the non-unitary operator $V(t)$ has a period $T$, and $M$ is a time-independent operator.
In addition, $U(t)$ is the unique fundamental matrix satisfying $U(0) = I$
because $U(t)$ is the time-evolution operator.
In contrast  to the Hermite Hamiltonian case discussed in reference \cite{Moore1990a},
the $T$-periodic matrix $V(t)$ is non-unitary, and $M$ is non-Hermitian.
Because of the non-Hermitian property of the Hamiltonian,
we here introduce the non-unitary time-evolution operator of the periodic `adjoint' Hamiltonian $H^\dagger (t)$
\cite{Choutri2002}.
The operator is written as $\widetilde{U}(t)$, and
we decompose $\widetilde{U}(t)$, in a similar way to $U(t)$, 
as $\widetilde{U}(t) = \widetilde{V}(t) \rme^{\rmi \widetilde{M} t}$.
Choutri \textit{et al.} \cite{Choutri2002} have shown that
the non-adiabatic geometrical phase is given by
\begin{eqnarray}
\gamma_\alpha = \int_0^T \langle \widetilde{\phi}_\alpha (0) | \rmi \widetilde{V}^\dagger \dot{V} 
| \phi_\alpha(0) \rangle \rmd t,
\end{eqnarray}
where $|\phi_\alpha(0)\rangle$ are the right eigenvectors of $U(T)$,
and $| \widetilde{\phi}_\alpha (0) \rangle$ are those of $\widetilde{U}(T)$.
Each $|\phi_\alpha(0)\rangle$ is also an eigenvector of $M$,
because $U(T) = \rme^{\rmi M T}$.
We here note that $\langle \widetilde{\phi}_\alpha (0) |$ is 
not the conjugate of $| \phi_\alpha (0) \rangle$,
but that of $| \widetilde{\phi}_\alpha (0) \rangle$.
Finally, the pump current is evaluated from
\begin{eqnarray}
J_\mathrm{pump} = \frac{1}{T} \frac{\partial}{\partial \chi} \gamma_\alpha
= \frac{\omega}{2 \pi} \frac{\partial}{\partial \chi} \gamma_\alpha.
\label{eq_calc_pump}
\end{eqnarray}
However, it is difficult to obtain $V(t)$ and $M$ directly in general;
some additional analytical treatments are needed
in order to calculate the geometrical phase $\gamma_\alpha$.

\subsection{Geometrical phase in terms of the Shirley's fundamental matrix}

There are two fundamental matrices
of particular importance for calculating the geometrical phase \cite{Moore1990a}.
The first is the time-evolution matrix $U(t)$ which is discussed above.
The second is the fundamental matrix $F(t) = P(t) \rme^{\rmi Q t}$, which is discussed by Shirley 
\cite{Shirley1965}.
The fundamental matrix $F(t)$ has suitable properties for the Fourier analysis,
which enables us to calculate $V(t)$ and $M$ explicitly.
In the Hermitian case discussed in references \cite{Shirley1965,Moore1990a},
the matrix $Q$ is real and diagonal in some convenient basis.
In the non-Hermitian case, the matrix $Q$ is not diagonalized in general,
and hence all discussions would be modified by using the Jordan canonical form
(but we do not use the Jordan canonical form explicitly
because the matrix $Q$ can be diagonalized in our case, as shown later.)

Henceforth, we will sometimes omit the time dependence of operators (matrices)
for notational simplicity (such as $U \equiv U(t)$).
Since $U$ and $F$ are both fundamental matrices of $H$,
there exists a constant invertible matrix $X$ with $U = F X$ \cite{Moore1990a}.
Because $U(0) = I$, we have $U(0) = F(0) X = I$.
Hence, we obtain $X = F(0)^{-1}$ and 
\begin{eqnarray}
U(t) = F(t) F(0)^{-1} = P(t) \rme^{\rmi Q t} P(0)^{-1}
= P(t) P(0)^{-1} \rme^{\rmi P(0) Q P(0)^{-1} t}.
\end{eqnarray}
We can here make the identifications
$V(t) = P(t) P(0)^{-1}$ and $M = P(0) Q P(0)^{-1}$.
When we set the eigenvectors of $Q$ as $| \alpha \rangle$,
the eigenvectors of $M$, $|\phi_\alpha(0) \rangle$,
are related to $| \alpha \rangle$ by $|\phi_\alpha(0) \rangle = P(0) | \alpha \rangle$.
We perform the same calculation for the time-evolution operator $\widetilde{U}(t)$,
and obtain $\widetilde{V}(t) = \widetilde{P}(t) \widetilde{P}(0)^{-1}$ and 
$\langle \widetilde{\phi}_\alpha(0) | = \langle \widetilde{\alpha} | \widetilde{P}(0)^{\dagger}$.
The non-adiabatic geometrical phase is therefore rewritten as
\begin{eqnarray}
\gamma_\alpha &= \int_0^T \langle \widetilde{\phi}_\alpha (0) | 
\rmi \left[ \widetilde{P}(0)^\dagger \right]^{-1}  \widetilde{P}(t)^\dagger
\dot{P}(t) P(0)^{-1}
| \phi_\alpha(0) \rangle \rmd t \nonumber \\
&= \int_0^T \langle \widetilde{\alpha} | 
\rmi \widetilde{P}(t)^\dagger \dot{P}(t)
| \alpha \rangle \rmd t.
\end{eqnarray}

\subsection{The Floquet Hamiltonian}

So far, we have shown that it is possible to write the non-adiabatic geometrical phase
in terms of the fundamental matrix $F(t) = P(t) \rme^{\rmi Q t}$ even in the non-Hermitian case.
The next step is the calculation of the explicit form of $P(t)$.
For that purpose, Shirley has introduced the Floquet Hamiltonian $H_\mathrm{F}$ as follows.
We introduce the Fourier expansion of the Hamiltonian $H$ as
\begin{eqnarray}
H = \sum_{n=-\infty}^{\infty} H^{(n)} \rme^{\rmi n \omega t},
\end{eqnarray}
where the Fourier components are defined by
\begin{eqnarray}
H^{(n)} = (1/T) \int_0^T H \rme^{-\rmi n \omega t} \rmd t
\label{eq_fourier_right}.
\end{eqnarray}
The Floquet Hamiltonian is defined by the matrix elements \cite{Shirley1965}
\begin{eqnarray}
\left\langle \alpha', n \right| H_\mathrm{F} \left| \beta, m \right\rangle
= H_{\alpha' \beta}^{(n-m)} + n \omega \delta_{\alpha'\beta} \delta_{nm},
\end{eqnarray}
where $| \alpha, n \rangle$ is an orthonormal basis for the matrix representation of $H_\mathrm{F}$.
The index $\alpha$ represents a spatial part and the index $n$ merely represents a Fourier components.
In our case, we have only two spatial parts (as we will see it later), 
and denote them by $+$ or $-$, i.e., $\alpha, \beta \in \{ +, - \}$.
The $| \alpha, n \rangle$ is called a `Floquet state' \cite{Shirley1965}.
Note that $H$ is non-Hermitian, so $\langle \alpha', n |$ is not a conjugate state of $| \alpha, n \rangle$,
but the corresponding left orthonormal basis;
we added the prime on $\alpha$ in order to clarify this fact.

\subsection{The relationship between the Floquet Hamiltonian and the Shirley fundamental matrix}

Like for the case with the Hamiltonian $H$, we define the Fourier expansion of $P(t)$ as
\begin{eqnarray}
P = \sum_{n=-\infty}^{\infty} P^{(n)} \rme^{\rmi n \omega t},
\end{eqnarray}
Shirley \cite{Shirley1965} has shown that the Fourier components of $P$ have a simple relationship
between the Floquet Hamiltonian in the Hermitian case.
Let $H_\mathrm{F}$ have eigenvectors $| \varepsilon_{\alpha,n} \rangle$ and eigenvalues $\varepsilon_{\alpha,n}$.
In the Hermitian case,
the matrix elements of $Q$ and the Fourier coefficients of $P$ are given by 
$Q_{\alpha \beta} = - \varepsilon_{\alpha, 0} \delta_{\alpha \beta}$
and $P_{\alpha' \beta}^{(n)} = \langle \alpha',n | \varepsilon_{\beta,0} \rangle$ \cite{Shirley1965}.

In the non-Hermitian case, the matrix $Q$ is not diagonalized in general, so that we cannot 
continue the same discussion as for the Hermitian case.
However, when the eigenvectors of the Floquet Hamiltonian, $| \varepsilon_{\alpha,n} \rangle$,
are not degenerate,
the matrix $Q$ is diagonalized and we obtain the same expression for $Q_{\alpha \beta}$ and $P_{\alpha \beta}^{(n)}$
as for the Hermitian case.
Actually, in our case, the eigenvectors of the Floquet Hamiltonian is not degenerate, as shown in section 5
\cite{note_degenerate}.

In order to derive the Fourier expression of $\widetilde{P}^\dagger$,
we introduce another definition of the Fourier transformation
\begin{eqnarray}
\widetilde{H}^{(n)} = (1/T) \int_0^T H \rme^{\rmi n \omega t} \rmd t
\label{eq_fourier_left},
\end{eqnarray}
and expend $\widetilde{P}^\dagger$ as
\begin{eqnarray}
\widetilde{P}^\dagger = \sum_{n = -\infty}^\infty \widetilde{P}^{(n)\dagger} \rme^{-\rmi n \omega t}.
\end{eqnarray}
By using the definition of the Fourier components in equation~\eref{eq_fourier_left},
other Floquet Hamiltonian and Floquet states are introduced:
\begin{eqnarray}
\left\langle \widetilde{\alpha, n} \right| \widetilde{H}_\mathrm{F} \left| \widetilde{\beta', m} \right\rangle
= \widetilde{H}_{\alpha \beta'}^{(n-m)} 
- n \omega \delta_{\alpha \beta'} \delta_{nm}.
\end{eqnarray}
The matrix elements of $[\widetilde{P}^{(n)\dagger}]_{\alpha \beta'}$ are given by
$\langle \widetilde{\varepsilon}_{\alpha,0} | \widetilde{\beta', n}\rangle$,
where $\langle \widetilde{\varepsilon}_{\alpha,0} |$ are the `left' eigenvectors of the Floquet Hamiltonian
$\widetilde{H}_\mathrm{F}$.
(Unlike for the case of $H_\mathrm{F}$, we add the prime on $\alpha$ for the right eigenvectors of the
Floquet Hamiltonian $\widetilde{H}_\mathrm{F}$.)
From the orthogonality relations $\int_0^{T} \exp[\rmi (n-m) \omega t] \rmd t = T \delta_{nm}$,
we obtain
\begin{eqnarray}
\rmi \widetilde{P}^\dagger \dot{P} = \sum_{n,m=-\infty}^{\infty} (- n \omega)
\widetilde{P}^{(m)\dagger} P^{(n)}
\rme^{\rmi (n-m) \omega t},
\end{eqnarray}
and finally, the geometrical phase is expressed by means of the Fourier components of $P$ and $\widetilde{P}^\dagger$
as follows:
\begin{eqnarray}
\gamma_\alpha = \left\langle 
\widetilde{\alpha} \left| 
\sum_{n=-\infty}^\infty (-2 \pi n) \widetilde{P}^{(n) \dagger} P^{(n)}
\right| \alpha
\right\rangle
\label{eq_phase_fourier}.
\end{eqnarray}

\section{Perturbation calculation for pump flux}

\subsection{Perturbation for the Floquet eigenvectors}

\begin{table}
\caption{\label{table_floquet} A part of the Floquet Hamiltonian $H_\mathrm{F}$.
When $R = 0$, the Floquet Hamiltonian is block diagonalized.}
\footnotesize
\begin{tabular*}{\textwidth}{@{}c|cccccc}
\br
 $\,$ & \scriptsize $| +, n+1 \rangle$ &  \scriptsize $| -, n+1 \rangle$ &  \scriptsize $|+, n \rangle$ & 
 \scriptsize $| -, n \rangle$ &  \scriptsize $| +, n-1 \rangle$ &  \scriptsize $| -, n-1 \rangle$\\
\mr
 \scriptsize $\langle +', n+1 |$                         & 
 \scriptsize $-2\rmi b + (n+1)\omega$                    &  \scriptsize $a(\rmi  + \rmi \rme^{\rmi \chi})$ &  
 \scriptsize $\frac{1}{2}(-\rmi R - R)$                  &  \scriptsize $0$ &  
 \scriptsize $0$                                         &  \scriptsize $0$\\ 
 \scriptsize $\langle -', n+1 |$                         & 
 \scriptsize $b(\rmi + \rmi \rme^{-\rmi \chi})$           &  \scriptsize $-2\rmi a + (n+1) \omega$ & 
 \scriptsize $\frac{1}{2} (\rmi R + R \rme^{-\rmi \chi})$ &  \scriptsize $0$ & 
 \scriptsize $0$                                         &  \scriptsize $0$\\
 \scriptsize $\langle +', n |$ & 
 \scriptsize $\frac{1}{2} (-\rmi R + R)$ &  \scriptsize $0$ & 
 \scriptsize $-2\rmi b + n\omega$ &  \scriptsize $a(\rmi + \rmi \rme^{\rmi \chi})$ & 
 \scriptsize $\frac{1}{2} (-\rmi R - R)$ &  \scriptsize $0$ \\
 \scriptsize $\langle -', n |$ & 
 \scriptsize $\frac{1}{2}(\rmi R - R \rme^{-\rmi \chi})$ &  \scriptsize $0$ & 
 \scriptsize $b(\rmi + \rmi \rme^{-\rmi \chi})$ &  \scriptsize $-2\rmi a + n \omega$ & 
 \scriptsize $\frac{1}{2}(\rmi R + R \rme^{-\rmi \chi})$ &  \scriptsize $0$ \\
 \scriptsize $\langle +', n-1 |$ & 
 \scriptsize $0$ &  \scriptsize $0$ & 
 \scriptsize $\frac{1}{2}(-\rmi R + R)$ &  \scriptsize $0$ & 
 \scriptsize $-2\rmi b + (n-1)\omega$ &  \scriptsize $a(\rmi + \rmi \rme^{\rmi \chi})$\\
 \scriptsize $\langle -', n-1 |$ & 
 \scriptsize $0$ &  \scriptsize $0$ & 
 \scriptsize $\frac{1}{2}(\rmi R - R \rme^{-\rmi \chi})$ &  \scriptsize $0$ & 
 \scriptsize $b(\rmi + \rmi \rme^{-\rmi \chi})$ &  \scriptsize $-2\rmi a + (n-1) \omega$\\
\br
\end{tabular*}
\end{table}

In order to calculate the Fourier components of $P$ and $\widetilde{P}^\dagger$,
we need to calculate the explicit expression of the Floquet Hamiltonian
and its eigenvectors.
While it is easy to calculate such eigenvectors in the block diagonalized Floquet Hamiltonian
as in the case in \cite{Choutri2002},
the Floquet Hamiltonian in our case cannot be block diagonalized.
Table~\ref{table_floquet} gives a part of the Floquet Hamiltonian $H_\mathrm{F}$.
When $R=0$, it is easy to see that the Floquet Hamiltonian is block diagonalized.
Hence, we here use the standard perturbation techniques,
though a little modification is needed to calculate the perturbed states for the left eigenvectors of
the Floquet Hamiltonian $\widetilde{H}_\mathrm{F}$.

When $R = 0$, we can show that the infinite-dimensional Floquet Hamiltonian is 
block diagonalized with typical block
\begin{eqnarray}
H_0 = \left[\begin{array}{cc}
-2 \rmi b + n \omega & \rmi a + \rmi a \rme^{\rmi \chi} \\
\rmi b + \rmi b \rme^{- \rmi \chi} & -2 \rmi a + n \omega
\end{array}\right]
\label{eq_right_hamiltonian_0},
\end{eqnarray}
in the basis $\{| +, n \rangle, |-,n \rangle \}$.
We define the perturbative part $H_\mathrm{F}'$ as the rest of the original Floquet Hamiltonian with $R \neq 0$.

From the Hamiltonian \eref{eq_right_hamiltonian_0},
we obtain the eigenvalues
\begin{eqnarray}
\varepsilon_{\pm, n}^{(0)} = -\rmi b - \rmi a \pm \rmi \rme^{- \rmi \chi} 
\sqrt{ \rme^{\rmi \chi} ( \rme^{\rmi \chi} b + a)(b + \rme^{\rmi \chi} a)} 
+ n \omega,
\end{eqnarray}
and the corresponding right eigenvectors
\begin{eqnarray}
\fl
| \varepsilon_{\pm, n}^{(0)} \rangle = z [ - \rme^{\rmi \chi} b + \rme^{\rmi \chi} a 
\pm  
\sqrt{ \rme^{\rmi \chi} ( \rme^{\rmi \chi} b + a)(b + \rme^{\rmi \chi} a)} 
] \left| +, n \right\rangle 
+ z (1+ \rme^{\rmi \chi}) b \left|-,n \right\rangle
\label{eq_H0_eigenvectors}.
\end{eqnarray}
The left eigenvectors of $H_0$ are obtained from
\begin{eqnarray}
\fl
\langle \varepsilon_{\pm', n}^{(0)} | = z' [ - \rme^{- \rmi \chi} b + \rme^{-\rmi \chi} a 
\pm  
\sqrt{ \rme^{-\rmi \chi} ( \rme^{-\rmi \chi} b + a)(b + \rme^{-\rmi \chi} a)} 
] \left\langle +', n \right| \nonumber \\
+ z' (1+ \rme^{-\rmi \chi}) a  \left\langle -',n \right|.
\end{eqnarray}
$z$ and $z'$ are normalization constants satisfying 
$\langle \varepsilon_{\alpha', n}^{(0)} | \varepsilon_{\beta, m}^{(0)} \rangle = \delta_{\alpha' \beta} \delta_{nm}$.

From the usual perturbation theory, we obtain the first order correction for the state 
$|\varepsilon_{+, n}^{(0)} \rangle$ as
\begin{eqnarray}
\fl
| \varepsilon_{+, n}^{(1)} \rangle = 
\frac{\langle \varepsilon_{+', n+1}^{(0)} | H'_\mathrm{F} | \varepsilon_{+, n}^{(0)} \rangle }
{\varepsilon_{+,n}^{(0)} - \varepsilon_{+,n+1}^{(0)}}
| \varepsilon_{+, n+1}^{(0)} \rangle
+
\frac{\langle \varepsilon_{-', n+1}^{(0)} | H'_\mathrm{F} | \varepsilon_{+, n}^{(0)} \rangle }
{\varepsilon_{+,n}^{(0)} - \varepsilon_{-,n+1}^{(0)}}
| \varepsilon_{-, n+1}^{(0)} \rangle \nonumber \\
+
\frac{\langle \varepsilon_{+', n-1}^{(0)} | H'_\mathrm{F} | \varepsilon_{+, n}^{(0)} \rangle }
{\varepsilon_{+,n}^{(0)} - \varepsilon_{+,n-1}^{(0)}}
| \varepsilon_{+, n-1}^{(0)} \rangle
+
\frac{\langle \varepsilon_{-', n-1}^{(0)} | H'_\mathrm{F} | \varepsilon_{+, n}^{(0)} \rangle }
{\varepsilon_{+,n}^{(0)} - \varepsilon_{-,n-1}^{(0)}}
| \varepsilon_{-, n-1}^{(0)} \rangle.
\end{eqnarray}
The first order correction for $|\varepsilon_{-, n}^{(0)} \rangle$ is calculated in a similar manner.
The matrix elements of $P_{\alpha' \beta}^{(n)}$ are therefore
given by
\begin{eqnarray}
P_{\alpha' \beta}^{(n)} = \langle \alpha', n | 
\left( | \varepsilon_{\beta, 0}^{(0)} \rangle + | \varepsilon_{\beta, 0}^{(1)} \rangle \right)
\label{eq_P_right}.
\end{eqnarray}

For the Floquet Hamiltonian $\widetilde{H}_\mathrm{F}$,
we perform the similar calculation,
but in the present case,
we need the first order correction of the left eigenvectors $\langle \widetilde{\varepsilon}_{+, n}^{(0)}|$.
The Floquet Hamiltonian $\widetilde{H}_\mathrm{F}$ has different Floquet states to $H_\mathrm{F}$,
and the diagonal part of the matrix representation of $\widetilde{H}_\mathrm{F}$ is given by
\begin{eqnarray}
\begin{array}{c}
\langle \widetilde{+, n} | \widetilde{H}_\mathrm{F} | \widetilde{+', n} \rangle = - 2 \rmi b - n \omega,\\
\langle \widetilde{-, n} | \widetilde{H}_\mathrm{F} | \widetilde{-', n} \rangle = - 2 \rmi a - n \omega.
\end{array}
\end{eqnarray}
The other matrix elements of the Floquet Hamiltonian $\widetilde{H}_\mathrm{F}$ are the same as 
those of the Floquet Hamiltonian $H_\mathrm{F}$.
Hence, we need a slight modification for the case of $\widetilde{H}_\mathrm{F}$ as follows:
\begin{eqnarray}
\widetilde{H}_0 = \left[\begin{array}{cc}
-2 \rmi b - n \omega & \rmi a + \rmi a \rme^{\rmi \chi} \\
\rmi b + \rmi b \rme^{- \rmi \chi} & -2 \rmi a - n \omega
\end{array}\right],
\label{eq_left_hamiltonian_0}
\end{eqnarray}
\begin{eqnarray}
\widetilde{\varepsilon}_{\pm, n}^{(0)} = -\rmi b - \rmi a \pm \rmi \rme^{- \rmi \chi} 
\sqrt{ \rme^{\rmi \chi} ( \rme^{\rmi \chi} b + a)(b + \rme^{\rmi \chi} a)} 
- n \omega,
\end{eqnarray}
\begin{eqnarray}
\fl
\begin{array}{l}
| \widetilde{\varepsilon}_{\pm', n}^{(0)} \rangle = z [ - \rme^{\rmi \chi} b + \rme^{\rmi \chi} a 
\pm  
\sqrt{ \rme^{\rmi \chi} ( \rme^{\rmi \chi} b + a)(b + \rme^{\rmi \chi} a)} 
] | \widetilde{+', n} \rangle 
+ z (1+ \rme^{\rmi \chi}) b | \widetilde{-',n} \rangle, \\
\langle \widetilde{\varepsilon}_{\pm, n}^{(0)} | = z' [ - \rme^{- \rmi \chi} b + \rme^{-\rmi \chi} a 
\pm  
\sqrt{ \rme^{-\rmi \chi} ( \rme^{-\rmi \chi} b + a)(b + \rme^{-\rmi \chi} a)} 
] \langle \widetilde{+, n} | 
+ z' (1+ \rme^{-\rmi \chi}) a  \langle \widetilde{-,n} |.
\end{array}
\end{eqnarray}
The first order correction for the $\langle \widetilde{\varepsilon}_{+, n}^{(0)}|$ is calculated from
\begin{eqnarray}
\fl
\langle \widetilde{\varepsilon}_{+, n}^{(1)} | = 
\frac{\langle \widetilde{\varepsilon}_{+, n}^{(0)} | \widetilde{H}'_\mathrm{F} 
| \widetilde{\varepsilon}_{+', n+1}^{(0)} \rangle }
{\widetilde{\varepsilon}_{+,n}^{(0)} - \widetilde{\varepsilon}_{+,n+1}^{(0)}}
\langle \widetilde{\varepsilon}_{+, n+1}^{(0)} |
+
\frac{\langle \widetilde{\varepsilon}_{+, n}^{(0)} | \widetilde{H}'_\mathrm{F} 
| \widetilde{\varepsilon}_{-', n+1}^{(0)} \rangle }
{\widetilde{\varepsilon}_{+,n}^{(0)} - \widetilde{\varepsilon}_{-,n+1}^{(0)}}
\langle \widetilde{\varepsilon}_{-, n+1}^{(0)} | \nonumber \\
+
\frac{\langle \widetilde{\varepsilon}_{+, n}^{(0)} | \widetilde{H}'_\mathrm{F} 
| \widetilde{\varepsilon}_{+', n-1}^{(0)} \rangle }
{\widetilde{\varepsilon}_{+,n}^{(0)} - \widetilde{\varepsilon}_{+,n-1}^{(0)}}
\langle \widetilde{\varepsilon}_{+, n-1}^{(0)} |
+
\frac{\langle \widetilde{\varepsilon}_{+, n}^{(0)} | \widetilde{H}'_\mathrm{F} 
| \widetilde{\varepsilon}_{-', n-1}^{(0)} \rangle }
{\widetilde{\varepsilon}_{+,n}^{(0)} - \widetilde{\varepsilon}_{-,n-1}^{(0)}}
\langle \widetilde{\varepsilon}_{-, n-1}^{(0)} |,
\end{eqnarray}
and the matrix elements of $\widetilde{P}^{(n) \dagger}$ are given by
\begin{eqnarray}
[\widetilde{P}^{(n) \dagger}]_{\alpha \beta'} = 
\left( \langle \widetilde{\varepsilon}_{\alpha, 0}^{(0)} | 
+ \langle \widetilde{\varepsilon}_{\alpha, 0}^{(1)} | \right)
| \widetilde{\beta', n} \rangle
\label{eq_P_left}.
\end{eqnarray}

\subsection{Explicit expressions for the pump current and some discussions}

From equations~\eref{eq_phase_fourier}, \eref{eq_P_right} and \eref{eq_P_left},
we finally obtain the geometrical phase $\gamma_{+}$ and $\gamma_{-}$ after tedious calculation.
We here note that there are two cyclic states corresponding to the spatial part $+$ and $-$.
While there are two cyclic states, we expect that the pump flux to be given by $\gamma_{+}$.
The reason is as follows.
Since the coefficients of the states $| +, n\rangle$ and $| -,n\rangle$ are expected to be related to
the probability $\mathbf{p}(t)$,
these coefficients should be positive when $\chi = 0$.
From the definition of the kinetic rates, we have positive parameters $a$ and $b$ ($a,b > 0$),
so that we expect that the eigenvector with the $+$ spatial part gives a suitable geometrical phase
for the classical pump current.
(For the eigenvector with the $-$ spatial part, one of the coefficients of $| +, n\rangle$ and $| -,n\rangle$ 
in equation \eref{eq_H0_eigenvectors} becomes negative when $\chi = 0$.)
When $\gamma_{+}$ is adopted,
we obtain the following pump current using equation~\eref{eq_calc_pump}:
\begin{eqnarray}
J_\textrm{pump} = \frac{b}{4a} R^2 \frac{\omega}{ 4(a+b)^2 + \omega^2},
\label{eq_current_result}
\end{eqnarray}
and in particular, for the adiabatic case ($\omega \ll 1$),
\begin{eqnarray}
J_\textrm{pump}^\textrm{\scriptsize(adiabatic)} \simeq \frac{b}{16 a (a+b)^2} R^2 \omega.
\label{eq_current_result_adiabatic}
\end{eqnarray}
We note that the geometrical phase in equation~\eref{eq_phase_fourier} consists of 
the two eigenvectors $| \alpha \rangle$ and $\langle \widetilde{\alpha} |$,
and hence the first-order correction in $R$ of the Floquet eigenvectors gives the second order in $R$
of equations~\eref{eq_current_result} and \eref{eq_current_result_adiabatic}.
This is also consistent with the result in \cite{Sinitsyn2007} in the small $R$ regime \cite{note_R}.
In addition, the dependence of the pump current on the frequency $\omega$ is consistent with the pump current 
in \cite{Sinitsyn2007}.
Hence, we consider that the eigenvector with the $+$ spatial part gives a suitable pump current in our formalism.

For the case with $\omega \gg 1$,
equation~\eref{eq_current_result} shows the power-law decay $\sim 1/\omega$,
which is expected by the other discussions for the pump current \cite{Liu1990,Astumian2003,Jain2007}.
Furthermore, we have a stochastic-resonance-like behavior with respect to the frequency $\omega$:
there is a single maximum of the pump current at $\omega_\textrm{r} = 2 (a+b)$.
This behavior is the same as those obtained experimentally and theoretically \cite{Liu1990,Astumian2003,Jain2007}.
We therefore conclude that 
the discussion based on the geometrical phase for the pump current gives the pump current adequately.

\section{Concluding remarks}

In the present paper,
we calculate the pump current of a classical chemical system
by means of the geometrical phase.
First, we extend the Floquet theory to the non-Hermitian case,
which is applicable for the Floquet Hamiltonian with non-degenerate eigenstates\cite{note_degenerate}.
Secondly, we obtain an explicit expression for the pump current with the aid of a perturbation technique.
While we analysed only a specific case, our formulation is applicable for other examples \cite{note}.
One of the advantages compared to the known techniques
is that our formulation gives a formal recipe for the calculation of the pump current.
Combining the Floquet theory and the perturbation techniques,
the explicit dependence of the pump current on the frequency $\omega$ is derived.
In addition, it is possible to calculate the fluctuations of the pumping effects
(higher cumulants) using the derivatives of the geometrical phase \cite{Sinitsyn2007}.

The derived pump current is consistent with the previous results which have been obtained by the other approaches:
the pump current depends on the frequency of the perturbative kinetic rate linearly at low frequency regime.
In addition, there is a power-law decay of the pump current with respect to the frequency in the high frequency regime.
Note that we use the perturbation method in our calculation, so that the final result is provided
in the limit of weak perturbations.
In order to get quantitatively good results even in the large $R$ cases,
it would be needed to consider higher orders of $R$ or to use other analytical treatments.
However, the dependence of the pump current on the frequency $\omega$ is consistent with the previous results,
and we believe that our formulation would be useful for further discussions about the geometrical phase
and the pump current.

These treatments based on the geometrical phase would give
a unifying theory for such pumping effects.
In addition, the analogy with such geometrical phase in the quantum mechanics
might give us a deeper understanding of the classical system.
We expect that the physical interpretation of the eigenvectors of the Floquet Hamiltonian
might give us more information about the current,
and it would be possible to obtain deeper understanding of the stochastic resonance-like behavior
by considering it with the aid of the geometrical interpretation.




\begin{thebibliography}{99}
\bibitem{note_degenerate} When the Floquet eigenvectors are degenerate,
another analytical approach would be needed; this is future work.
\bibitem{note_R} When the second order correction of the Floquet eigenstates is considered,
the third and forth order in $R$ would arise in the pump current.
\end{thebibliography}
\end{document}